\documentclass[1p]{elsarticle}

\usepackage{lineno,hyperref}

\usepackage{amsmath,amssymb}
\usepackage{graphicx, color, psfrag, subfigure, float}
\usepackage{pstricks,pst-node,pst-plot,pst-circ,moredefs}
\usepackage{dcolumn}
\usepackage{bm}
\usepackage{multirow}
\usepackage{threeparttable}
\usepackage{dcolumn}
\usepackage{subfigure}
\usepackage{subfigmat}

\journal{Journal of International Heat and Fluid Flow}

\begin{document}

\begin{frontmatter}

\title{Unsteady High-Lift Mechanisms from Heaving Flat Plate Simulations}

\author{Jennifer A. Franck, Kenneth S. Breuer}
\address{School of Engineering, Brown University, Providence, RI}





\begin{abstract}
Flapping animal flight is often modeled as a combined pitching and heaving motion in order to investigate the unsteady flow structures and resulting forces that could augment the animal's lift and propulsive capabilities.  This work isolates the heaving motion of flapping flight in order to numerically investigate the flow physics at a Reynolds number of 40,000, a regime typical for large birds and bats and challenging to simulate due to the added complexity of laminar to turbulent transition in which boundary layer separation and reattachment are traditionally more difficult to predict. Periodic heaving of a thin flat plate at fixed angles of attacks of 1, 5, 9, 13, and 18 degrees are simulated using a large-eddy simulation (LES).  The heaving motion significantly increases the average lift compared with the steady flow, and also surpasses the quasi-steady predictions due to the formation of a leading edge vortex (LEV) that persists well into the static stall region. The progression of the high-lift mechanisms throughout the heaving cycle is presented over the range of angles of attack. Lift enhancement compared with the equivalent steady state flow was found to be up to 17\% greater, and up to 24\% greater than that predicted by a quasi-steady analysis.  For the range of kinematics explored it is found that maximum lift enhancement occurs at an angle of attack of 13 degrees, with a maximum lift coefficient of 2.1, a mean lift coefficient of 1.04. 
\end{abstract}


\end{frontmatter}

\section{Introduction}

There has been much interest in the aerodynamics community surrounding the flapping flight of insects, birds, and bats, due to the high-lift mechanisms they exploit during a typical wing-beat such as delayed stall, vortex/wake recapture, or clap and fling. Naturally, the question arises if these same mechanisms can be duplicated in man-made aircraft such as micro-air vehicles (MAV), which requires detailed investigations of the kinematics, mechanics, and control surrounding a flapping wing. The engineering community has traditionally been interested in the rigid motion of airfoils and the unsteady effects of dynamic stall for rotary-wing aircraft \cite{McCroskey1982} applications, and more recently, wind turbines \cite{Leishman2002}, both of which are within a Reynolds number regime of $10^6$.  However, much of the recent computational and experimental work has been focused on the unsteady flapping motion of insect flight, who typically fly at Reynolds numbers of $O(10^2-10^3)$. This work is focused in the intermediate Reynolds number regime of $O(10^3-10^5)$, of which there have been fewer publications on flapping flight and the aerodynamics of larger vertebrates, including bats and birds.  Using a flat plate computational model, and a time-resolved large-eddy simulation (LES), this work provides insight into the unsteady high-lift mechanisms of flapping flight at an intermediate Reynolds number of 50,000.

The kinematics and mechanics of flapping flight for insects such as clap and fling, wake capture, and delayed stall have been thoroughly investigated and reported, as summarized by Shyy et al. \cite{Shyy2010} and more recently by Chin and Lentink \cite{chin2016}.  Of particular interest is the delayed stall phenomena, in which a leading edge vortex (LEV) on the upper edge of the wing enhances lift throughout the downstroke, which has been shown experimentally \cite{Ellington1996, BirchBIO04} and computationally \cite{Liu1998, Wang2000}. The presence of LEVs and other large flow structures indicates that traditional quasi-steady flow analysis that relies on inviscid flow theory is not capturing the correct physics.  One such inviscid model is that of Theodorsen \cite{Theo34}, who developed a unsteady lift prediction in 1935 composed of quasi-steady, added mass, and circulatory parts. However despite is wide use, the Theodorsen model neglects any type of boundary layer separation such as the formation of LEVs or other large-scale structures.

Evidence of unsteady vortex structures at intermediate Reynolds numbers has also been shown in animal flight, such as the wind tunnel testing of bats \cite{Muijres2008}. A common experimental and computational model of flapping flight is the sinusoidal pitching and/or heaving motion of an airfoil geometry in which an LEV is generated, and the effects of lift and drag can be carefully documented. Many experiments have looked at symmetric airfoils in sinusoidal heaving motion \cite{Lee2004, Young2004}, or pitching motion \cite{Rival2008}, finding the boundary layer separation and formation of a LEV can greatly impact the lift generation. Experiments by Cleaver et. al. \cite{Cleaver2011} have looked at small amplitude heaving of a NACA0012 airfoil at post-stall angles of attack at $Re=10,000$ and also found an enhancement in lift due to the formation and convection of LEVs. They further deduced that lift increased with increasing heaving frequency and plunge velocity. Direct numerical simulations (DNS) were performed to capture the transitional nature of the flow at this Reynolds number, and provide detailed dynamics of the LEV.
For a controlled growth of a LEV, Ford and Babinsky \cite{Ford2013} investigated the unsteady vortex formation on an accelerating flat plate, finding that most of the bound circulation remained in the LEV. Also interested in vorticity transport, Panah et al \cite{Panah2015} calculated the circulation during the LEV development on a plunging foil at a modest Reyolds number of $10,000$. 

Water tunnel experiments by Ol et al. \cite{Ol2009} performed two plunging configurations of an SD7003 airfoil in freestream conditions at $Re=10,000-60,000$. Although unsteady vortices were clearly present, the attached-flow theory still managed to predict proper trends in lift force. Simulations were preformed with a 2D RANS model, and captured the formation of the LEV, but could not accurately capture the reattachment, which has been previously documented in RANS computations of unsteady flows \cite{Rumsey04}. In a followup paper, a thin flat plate is compared to the SD7003 airfoil, and it is found the geometry can influence the LEV formation and lift forces by promoting earlier separation \cite{Kang2013}, and experimental studies have by Rival et al \cite{Rival2014} have shown a similar trend in a plunging plate of various leading edge geometries.  Visbal performed large-eddy simulations at $Re=60,000$ of a plunging SD7003 airfoil at an angle of attack of $8^{\circ}$, providing a detailed analysis of the LEV flow structure and 3D effects \cite{Visbal2011}.  

This study was motivated by recent wind tunnel experiments have been performed by Curet et al on a self-excited flapper in a uniform flow \cite{Curet11, Curet2013} that demonstrate an increase in lift after transitioning from a stationary to flapping mode. The flapper model is composed of two rigid plates, a main body connected to a trailing-edge flap with a sailcloth, and is mounted at a positive angle of attack. The trailing edge flap is free to pitch with respect to the main body, and the main body is mounted on a cantilever beam and free to heave.  The experiments show that at low velocities the flapper remains a motionless single flat plate at a fixed angle of attack.  However above a critical freestream velocity $U^{*}$ an instability develops and the flapper begins to passively heave in a sinusoidal heaving motion, driven by the trailing flap motion. Upon its transition to the heaving mode, the flapper displays a large increase in average lift, which is dependent on the angle of attack as well as the ratio of $U_{\infty}/U^{*}$.  

This research is a followup investigation that provides a more detailed explanation of the lift enhancement documented in the experiments by Curet et al using a computational model of the flapper. The computations simulate the flow over the main body (flat plate) of the flapper in its stationary position and in its heaving position, and computing the lift and vortex dynamics and comparing it with quasi-steady predictions.  Although only the main body of the flapper is modeled in the computations, it is hypothesized that the increase in lift experienced by the flapper is its heaving mode is primarily due to the LEV formation on the main body, and that the trailing flap provides the instability that drives the sinusoidal heaving motion but does not contribute significantly to the lift enhancement.  

The Reynolds number matches the experiments at $40,000$, and the fixed angle of attack in the computations is varied from $1^{\circ}$ to $18^{\circ}$.  A large-eddy simulation (LES) in utilized for the computations due to its ability to resolve the time-dependent kinematics and dynamics of unsteady flow structures, such as the LEV, that develops during the heaving motion. Unlike the blunt leading edge airfoil geometries previously computed \cite{Visbal2011,Cleaver2011,Ol2009}, the boundary layer on the thin flat plate model separates at the sharp leading-edge, producing a LEV even at small angles of attack. Using a well-resolved boundary layer, the LES solver is expected to predict both separation and any subsequent reattachment of the separated shear layer, which is usually under predicted or not captured at all with RANS models.  With this heaving flat-plate model, we focus our attention on the two-dimensional effects of the flapping motion as opposed to any three-dimensional tip effects that may be present in the experiment. In particular we are interested in the leading edge separation and subsequent formation and convection of a dynamic stall vortex, and the effects on lift force for various plunging configurations.  

\section{Simulation Details}

\subsection{Governing Equations and Computational Methods}

An incompressible LES is used to perform the simulations. The governing equations are the filtered Navier-Stokes equations,

\begin{eqnarray}
\label{e:fNS}
\frac{\partial \bar{u}_i}{\partial t} + \frac{\partial \bar{u}_i \bar{u}_j}{\partial x_j} &=& -\frac{1}{\rho} \frac{\partial \bar{p}}{\partial x_j} + \nu \frac{\partial^2 \bar{u}_i}{\partial x_j \partial x_j} + f_{b_{i}} - \frac{\partial \tau_{ij} }{\partial x_j} \\
\frac{\partial \bar{u}_i}{ \partial x_i}  &=& 0, 
\end{eqnarray}

where overbar represents a low-pass spatially filtered quantity, $u_i$ are the three components of velocity, $p$ is pressure, $\nu$ is kinematic viscosity, and $\rho$ is density.  The sub-grid scale stresses are calculated with a constant Smagorinsky model, where

\begin{equation}
\label{e:sgs}
\frac{\partial \tau_{ij} }{\partial x_j} = -2 C_s^2 \Delta^2 |\bar{S}| \bar{S}_{ij}
\end{equation}

and the filtered rate of strain is

\begin{equation}
\label{e:strain}
\bar{S}_{ij} = \frac{1}{2}(\frac{\partial \bar{u}_i}{\partial x_j} + \frac{\partial \bar{u}_j}{\partial x_i}).
\end{equation}

For all simulations the Smagorinsky constant is $C_s=0.1$, chosen to be on the lower end of the acceptable range to minimize superfluous dissipation. The effect of changing the Smagorinsky constant is briefly discussed in the context of the mesh resolution in the following section. Rigid body motion is added by prescribing the appropriate body forces, $\mathbf{f_b}$, to the momentum equations creating a non-inertial frame of reference. The governing equations are solved using OpenFOAM libraries \cite{openfoam}, with a custom-built LES sub-grid scale model and additional of the non-intertial terms for the rigid body motion. The LES solver utilizes a second-order accurate finite-volume scheme using Gaussian integration and linear interpolation from cell centers to cell faces, which is standard in OpenFOAM solvers.  A pressure-implicit split-operator method is used to solve for the pressure, a second order backwards time-stepping routine is implemented, and a preconditioned conjugate gradient method solves the matrix equations. 

\subsection{Flow Configuration of Heaving Plate Simulations}
\label{s:mesh}

The static and heaving flow over a flat plate is modeled by a thin ellipse of aspect ratio 50. For the heaving flat plate model, the ellipse is prescribed an oscillatory motion in the vertical direction by

\begin{equation}
\label{e:vy}
v_{y}(t) = -U_{T} \sin(2 \pi f t) 
\end{equation}

which is equivalent to a body force of the form

\begin{equation}
\label{e:rigidmotion}
f_{b,y}(t) =  U_{T} f \cos(2 \pi f t). 
\end{equation}

Sinusoidal kinematics are chosen because they closely represent the motion of the self-excited flapper \cite{Curet2013} as well as many biological flyers.

The computational domain is 30 chord lengths in the streamwise direction, 15 chord lengths in the transverse direction, and 0.2 chord lengths in the spanwise direction. The velocity boundary condition at the inlet is 

\begin{eqnarray}
\label{e:bc}
u_x(t) = U_{T} \cos(\alpha_{rel}) \\
u_y(t) = U_{T} \sin(\alpha_{rel})
\end{eqnarray}

where $\alpha_{rel}$ is the relative angle of attack defined by 

\begin{equation}
\label{e:alpharel}
\alpha_{rel}(t) = \alpha + \tan^{-1}\frac{v_y(t)}{U_{\infty}}.
\end{equation}

At the outflow boundary the velocity has zero gradient. The flow has periodic boundary conditions in the spanwise direction to simulate an infinite-length wing.  In each simulation the computational time step is chosen such that the CFL number remains below 1. 

\begin{figure}[htp]
\centering  
\includegraphics[width=10cm]{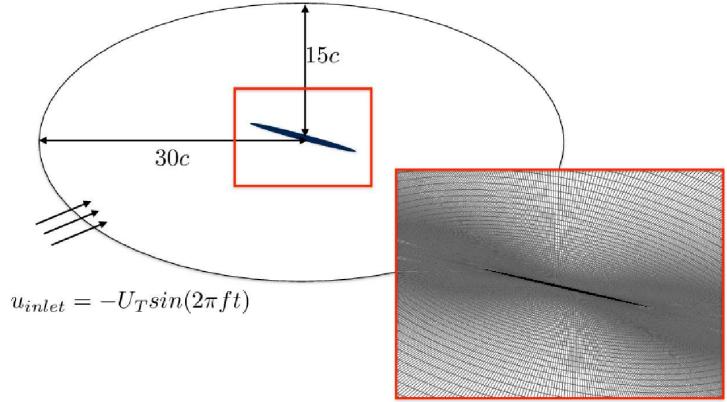}
\caption{Schematic of computational domain including the mesh surrounding the flat plate geometry, inclined to an angle of attack of $13^{\circ}$. The inlet boundary conditions vary with the sinusoidal heaving motion to satisfy the additional velocity imposed by the non-inertial reference frame.}
\label{f:mesh}
\end{figure} 

\begin{table}[htp]
 \begin{center}
 \begin{threeparttable}
  \caption{Mesh resolution study for the static flow at $\alpha=5^{\circ}$.}
 \label{t:mesh}
  \begin{tabular*}{0.95\textwidth}{@{\extracolsep{\fill}} c c c c c c c c c c}
  \hline
      Mesh &
   {$\overline{C_l}$} &
   {$\overline{C_d}$} &
   N &
   {$N_r$} &
   {$N_{\theta}$} &
   {$N_z$} &
   $\Delta r_{min}$ &
   $\Delta \theta_{min}$ & 
   $\Delta y^+$ \\ \hline
   mesh 1  & 0.562 & 0.052 & $2.09e^{6}$ & 190 & 344 & 32 & $9.75e^{-4}$ & $2.55e^{-4}$ & 0.98 \\
   mesh 2  & 0.467 & 0.046 & $1.83e^{6}$ & 192 & 368 & 26 & $3.42e^{-4}$ & $2.95e^{-4}$ & 0.33 \\
   mesh 3  & 0.468 & 0.046 & $3.53e^{6}$ & 240 & 460 & 32 & $2.68e^{-4}$ & $2.30e^{-4}$ & 0.26 \\
   mesh 4  & 0.469 & 0.047 & $6.04e^{6}$ & 288 & 552 & 38 & $2.34e^{-4}$ & $1.82e^{-4}$ & 0.22 \\ \hline
   \end{tabular*}
 \end{threeparttable}
 \end{center}
\end{table}

\begin{figure}[htp]
\centering  
\begin{subfigmatrix}{1}
\subfigure[pressure coefficient at the surface]{
\includegraphics[width=9cm]{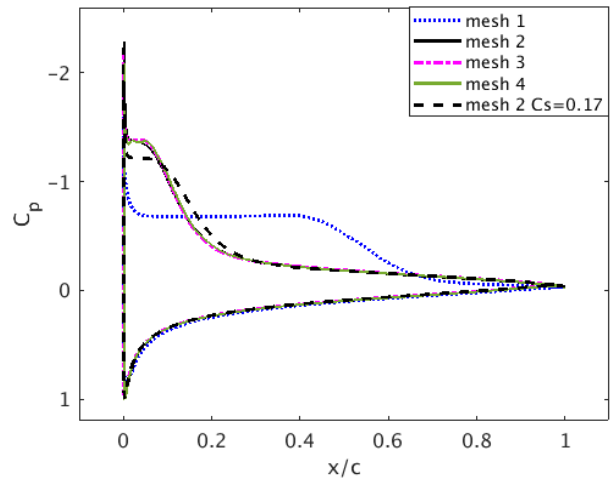}%
}
\subfigure[tangential velocity profiles along plate]{
\includegraphics[width=9cm]{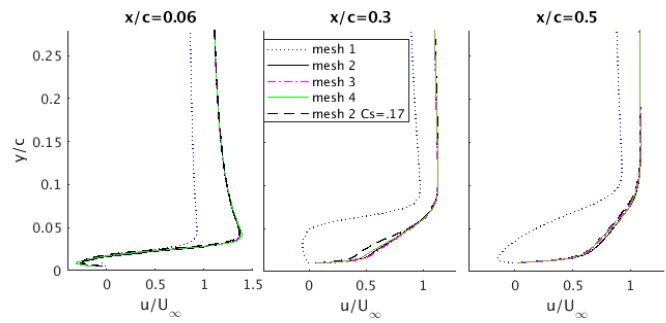}%
}
 \end{subfigmatrix}
\caption{Mesh resolution for static plate at $\alpha=5^{\circ}$.}
\label{f:meshres}
\end{figure} 

The grid is composed of 192 points in the radial direction, 368 points in the tangential direction, and 26 in the spanwise direction for a total of 1.83 million grid points, corresponding to mesh 2 in table \ref{t:mesh}. An example of the mesh in the proximity of the flat plate is shown in figure \ref{f:mesh}. A thorough mesh resolution study was performed for the static flow at $\alpha=5^{\circ}$ to evaluate the effect of the resolution on the flow. At $\alpha=5^{\circ}$ the flow exhibits a small separation bubble at the leading edge. Since the flow both separates and reattaches to the boundary layer the resolution study at this angle of attack will address the ability of the boundary layer to capture both of these phenomena.

Four meshes are presented in table \ref{t:mesh} which vary in tangential, radial, and spanwise resolution.  The number of points in each direction is reported, and due to various schemes for clustering points within the boundary layer, $\Delta r_{min}$, the first mesh point from the plate, and $\Delta \theta_{min}$, the smallest mesh point in the tangential direction, are also reported. The meshes are ranked by their $\Delta r_{min}$ resolution, from coarse to fine.  In each mesh $\Delta y^+<1$, and is computed on the mean flowfield at 75\% chord where the boundary layer is completely attached. 
  
The results of the mesh resolution study indicate that mesh 1 does not have enough radial resolution in the boundary layer to properly capture the reattachment of the boundary layer, despite having a $\Delta y^+<1$ and comparable spanwise and tangential resolution to the other meshes. The separated region at the leading edge of mesh 1 is 288\% larger than that of the other three meshes. This large discrepancy in the separation bubble size can be seen in figure \ref{f:meshres}(a) in terms of the pressure coefficient on the surface. In figure \ref{f:meshres}(b) the velocity profiles at $x\c=0.06$, $x/c=0.3$, and $x/c=0.5$ all show reverse flow for mesh 1, whereas the other meshes in the study show attached flow by $x/c=0.3$. The effect of increasing the Smagorinsky coefficient to 0.17 is also included in figure \ref{f:meshres}. The added dissipation by increasing $C_s$ increases the length of the separation bubble by 21\%, which is not as strong as an effect as the lower radial resolution in mesh 1. 

Meshes 2, 3, and 4, qualitatively agree on the separation bubble size, pressure profiles, and velocity profiles. Among these meshes, mesh 2 was chosen for the simulations due to its lower computational time, but still have an ability to properly predict the mean lift and drag coefficients, as well as the separation and reattachment dynamics of the boundary layer. 

\begin{figure}[htp]
\centering  
\begin{subfigmatrix}{1}
\subfigure[lab-fixed reference frame]{
\includegraphics[width=8cm]{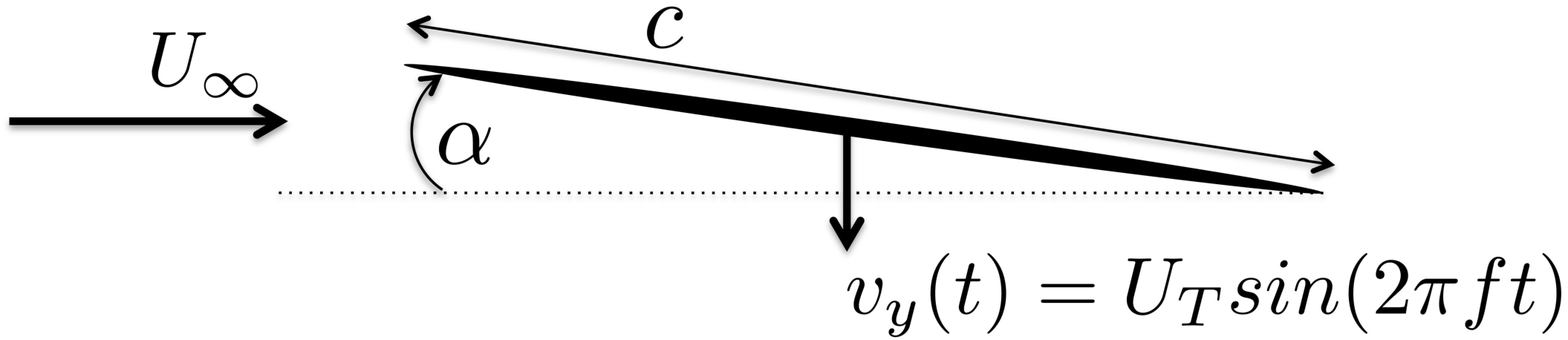}%
}
\subfigure[airfoil-fixed reference frame]{
\includegraphics[width=8cm]{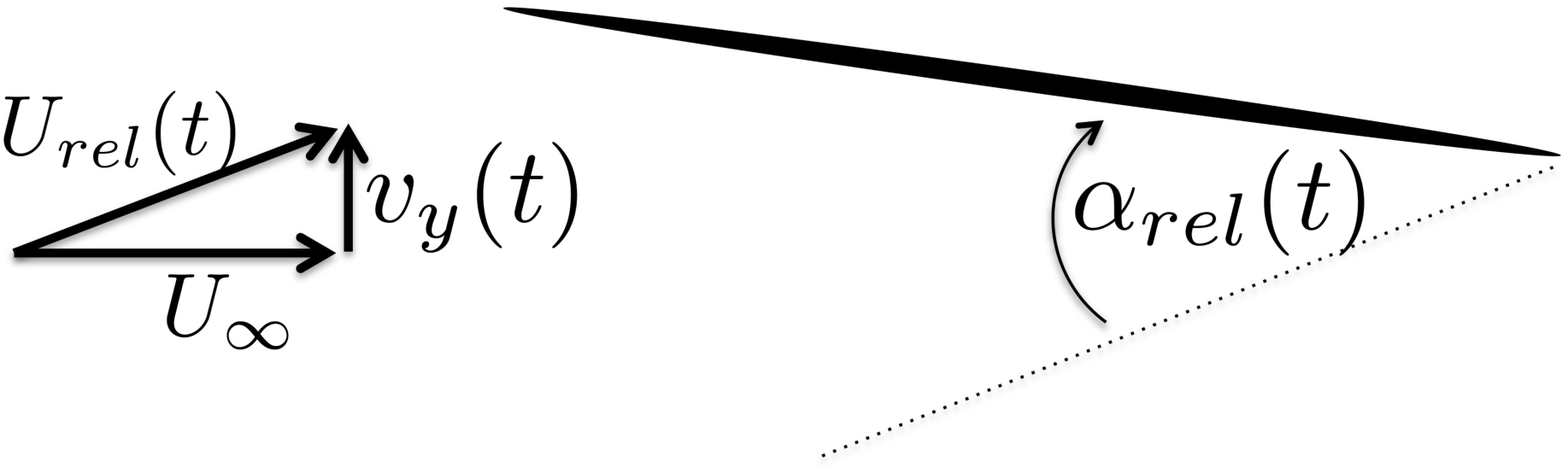}%
}
 \end{subfigmatrix}
\caption{Geometry and kinematics of the heaving airfoil.}
\label{f:kinematics}
\end{figure}

The non-dimensional simulation parameters are chosen to closely match the self-excited flapper experiments. The Reynolds number based on the chord $c$ and freestream velocity $U_{\infty}$ is $40,000$, the Strouhal number $St=fc/U_{\infty}$, is 0.08, and the maximum translational velocity $U_T/U_{\infty}$ is 0.12. It should be noted that the Strouhal number is slightly lower than many other flapping flight investigations in which $St=0.12-0.25$. Figure \ref{f:kinematics} summarizes the plate geometry, kinematics, and flow conditions.

Angles of attack between $1^{\circ}$ and $21^{\circ}$ are computed for the static, non-moving plate when $v_y=0$.  For the heaving plate simulations, five angles of attack are simulated at 1, 5, 9, 13, and 18 degrees. Simulations at a static angle of attack are allowed to develop into a steady state (from examination of the mean $C_l$) before the average lift coefficient is computed. Heaving plate simulations are initialized with the a static angle of attack flow field, and are run for a minimum of 5 heaving cycles, costing approximately 100 runtime hours on 64 processors using a SGI Altix ICE 8200 cluster.

\vspace{1cm}

\section{Static Plate Simulations}

\begin{figure}[htp]
\centering  
\includegraphics[width=3in]{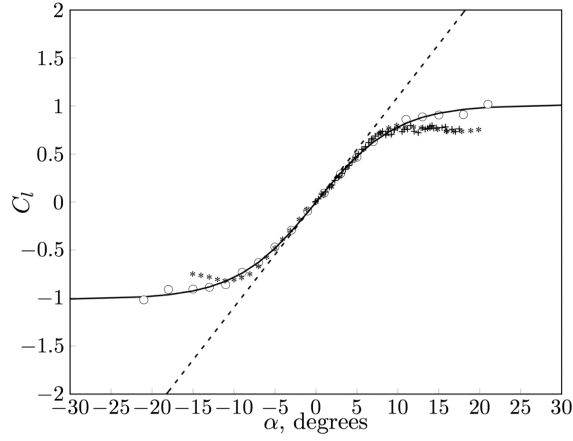}
\caption{Averaged lift coefficient $\bar{C_l}$ for various angles of attack $\alpha$ for a static plate. LES at $Re=40,000$ ($circ$), negative angles of attack are extrapolated due to airfoil symmetry. Thin airfoil theory (dashed), fitted LES data eqn. \ref{e:fit} (solid), $Re=80,000$ \cite{Pelletier00} (*), $Re=102,000$ \cite{Selig89} (+)}
\label{f:staticaoa}
\end{figure}

In order to provide a reference flow for the heaving plate simulations, a static airfoil in uniform flow is simulated at angles of attack ranging from $\alpha=1-21^{\circ}$ using the LES method described above. The mean lift coefficients were computed for each case and plotted in figure \ref{f:staticaoa}, where the negative angles of attack are extrapolated due to the symmetry of the airfoil, and classic thin airfoil theory of $C_l \approx 2 \pi \alpha$ is plotted for reference.  The average lift begins to deviate from thin airfoil theory at $\alpha=5^{\circ}$, and plateaus beginning at $\alpha=9^{\circ}$, indicating static stall.  Also plotted in figure \ref{f:staticaoa} are experimental data of lift on a thin flat plate at $Re=80,000$ \cite{Pelletier00}, and $Re=102,000$ \cite{Selig89}.  The current simulations agree well with the experimental data until the onset of static stall where the lift is slightly higher in the simulations. Since both sets of experiments are at a notably higher Reynolds number, the discrepancy could be a plausible Reynolds number effect. 

\begin{figure}[htp]
\centering  
\begin{subfigmatrix}{1}
\subfigure[]{
\includegraphics[width=8cm]{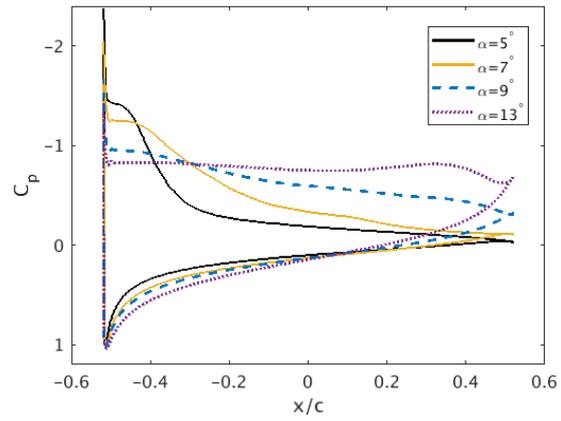}%
}
\subfigure[]{
\includegraphics[width=8cm]{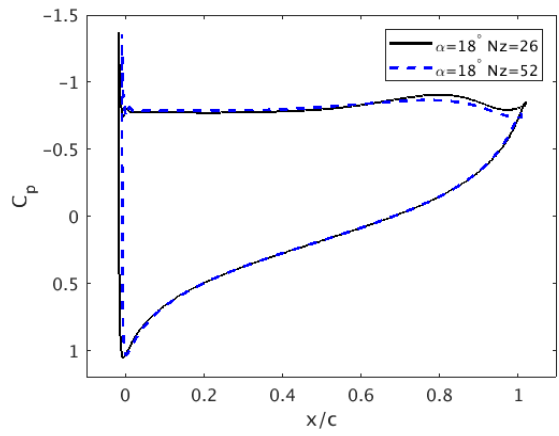}%
}
 \end{subfigmatrix}
\caption{Pressure coefficient for static angles of attack at $\alpha=5^{\circ}-18^{\circ}$, including the effect of spanwise resolution at $\alpha=18^{\circ}$.}
\label{f:cp}
\end{figure}

However, to further validate the static case within the stalled region, comparisons are made with previous LES and RANS simulations performed at $\alpha=18^{\circ}$ in a confined channel (walls at 1.5c above and below the plate) at $Re=20,000$, shown in table \ref{t:validation}. The resolution of the current LES compares favorably with that of the previous simulation in terms of number of cells in the boundary layer (on the non-inclined plate) and is more resolved in the tangential direction along the edge of the plate. Since it is significantly lower in spanwise resolution, another case with twice the number of points in span ($N_z=52$) is performed and also reported in table \ref{t:validation}. There is minimal difference between the mean lift and drag between the low and high spanwise resolution, and the frequency of oscillation in the lift signal remains the same.  There is a difference in terms of the magnitude of the fluctuations in the lift signal, with the higher resolution causing lower amplitude fluctuations. The difference causes minor changes in the pressure coefficient on the surface shown in figure \ref{f:cp}(b), and a slight shift downstream of the mean circulation bubble (not shown).

The current simulations have slightly lower lift and drag coefficients than the $Re=20,000$, thus lying between the higher Reynolds number experiments and the lower Reynolds number LES simulations.  In addition to a Reynolds number effect there are also differences due to the confinement within a channel, which may inflate the lift coefficient. However there is a strong qualitative agreement between the two sets of simulations in terms of pressure coefficient, streamlines, and shedding frequency. In addition to capturing the unsteady vortex shedding, the LES simulations are also better predictors of the lift and drag coefficients compared to the RANS simulations \cite{Breuer03}.  

%
\begin{table}[htp]
 \begin{center}
 \begin{threeparttable}
  \caption{Static flow simulations at $\alpha=18^{\circ}$ including resolution, lift, drag, vortex shedding frequency. Previous simulations \cite{Breuer03} are in a confined flow, and offer only a qualitative comparison.}
 \label{t:validation}
  \begin{tabular*}{0.99\textwidth}{@{\extracolsep{\fill}} c c c c c c c c c c}
  \hline
       &
       \small Re &
       \small N &
       \small cells in BL &
       \small $N_{\theta}$ &
       \small $N_z$ &       
      \small {$\overline{C_l}$} &
       \small $<C_l>$ &
   \small {$\overline{C_d}$} &
   \small $fc/U_{\infty}$ \\ \hline
   mesh 2, Nz=26          & 40k & $1.83e^{6}$ & 22 & 368 & 26 & 0.896 & 0.123 & 0.311 & 0.63 \\
   mesh 2, Nz=52          & 40k & $3.66e^{6}$ & 22 & 368 & 52 & 0.902 & 0.088 & 0.312 & 0.63 \\
   RANS-F \cite{Breuer03} & 20k & $5.2e^{4}$     & 20 & 256 & -  & 1.318 & -     & 0.316 & - \\
   LES-C \cite{Breuer03}  & 20k & $0.99e^{6}$    & 17 & 184 & 36 & 1.078 & -     & 0.365 & 0.63 \\
   LES-VF \cite{Breuer03} & 20k & $8.97e^{6}$    & 23 & 256 & 76 & 1.128 & -     & 0.380 & 0.64 \\ \hline
   \end{tabular*}
 \end{threeparttable}
 \end{center}
\end{table}

\vspace{1cm}

\begin{figure}[htp]
\subfigure[ $\alpha=5^{\circ}$]{
\includegraphics[width=0.3\textwidth]{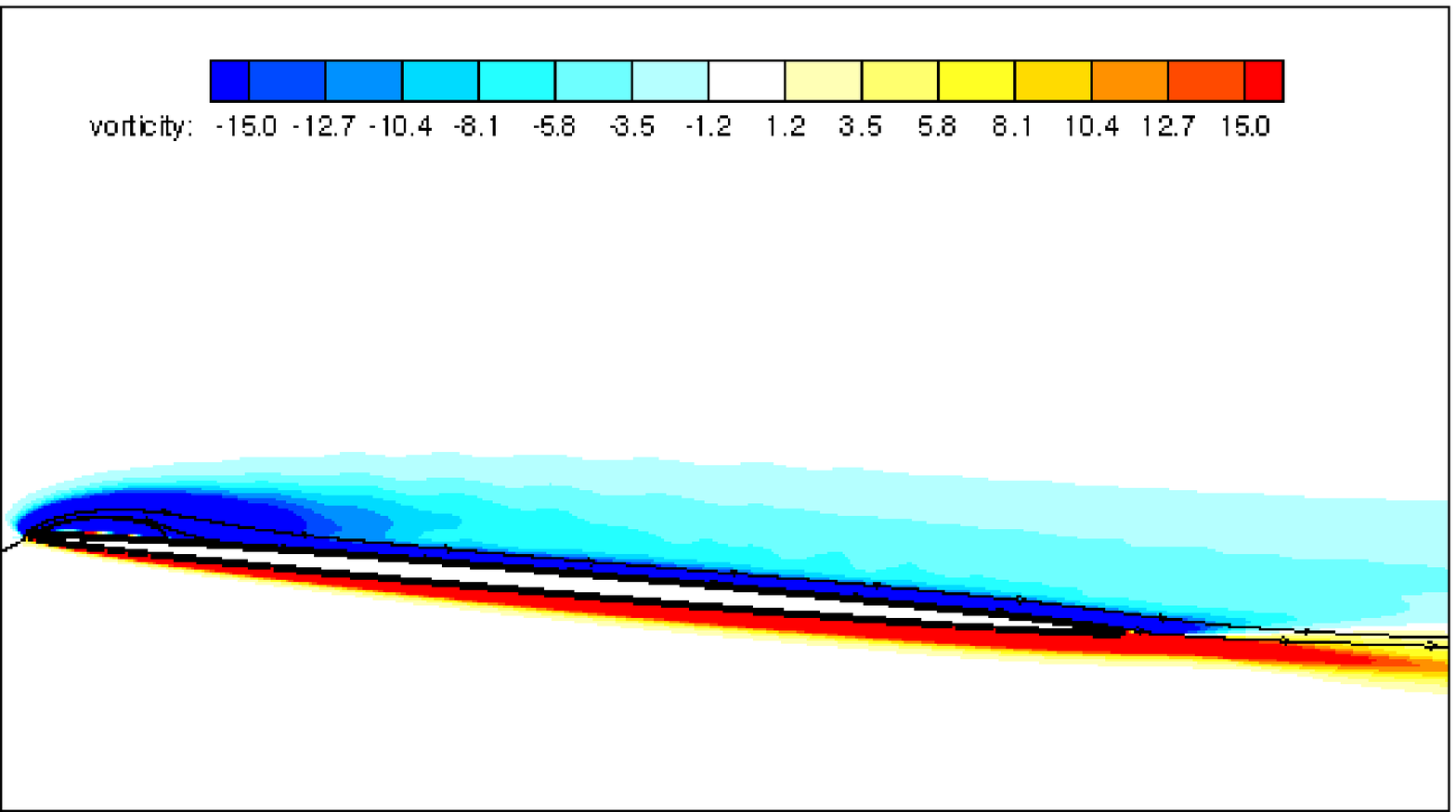}
}
\subfigure[ $\alpha=7^{\circ}$]{
\includegraphics[width=0.3\textwidth]{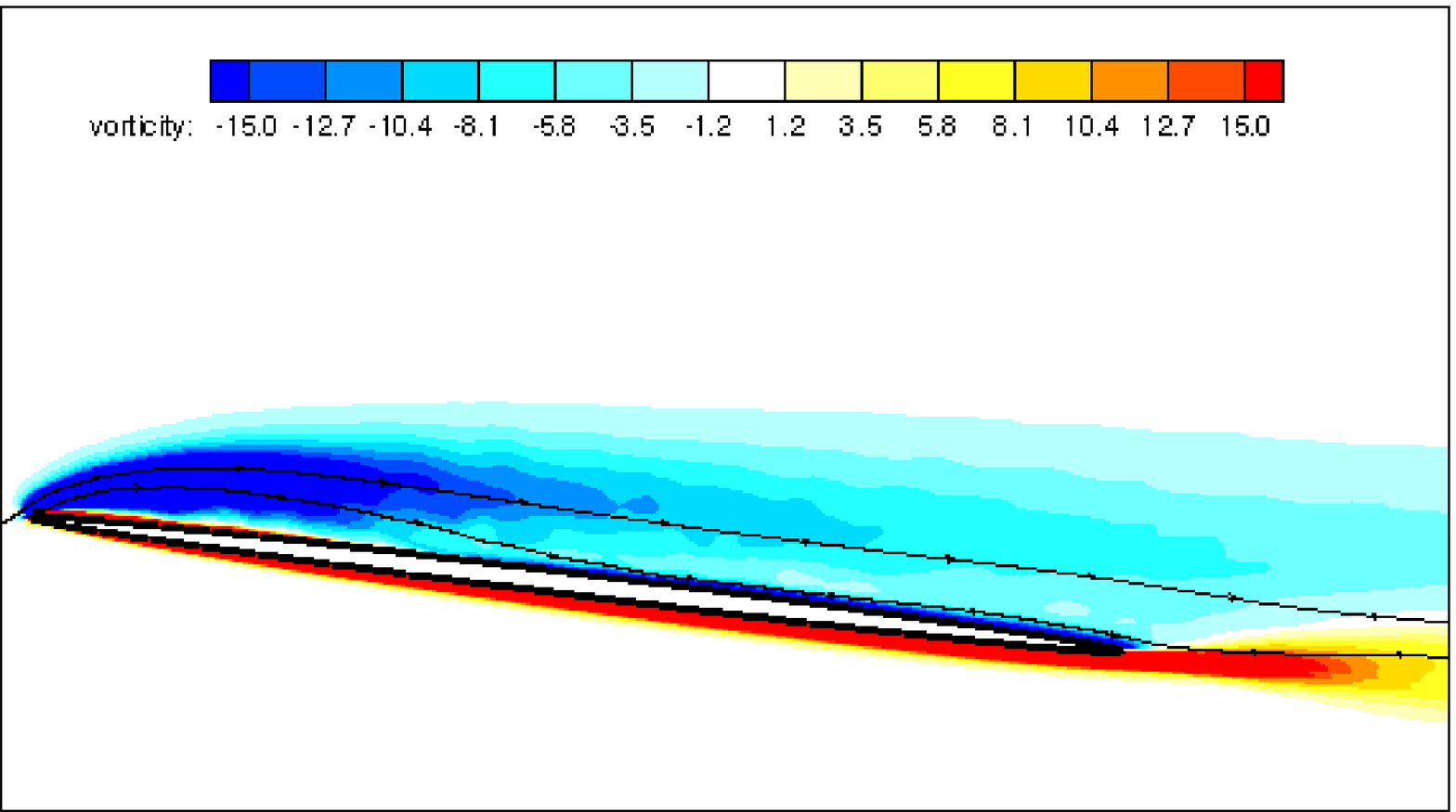}
}
\subfigure[ $\alpha=9^{\circ}$]{
\includegraphics[width=0.3\textwidth]{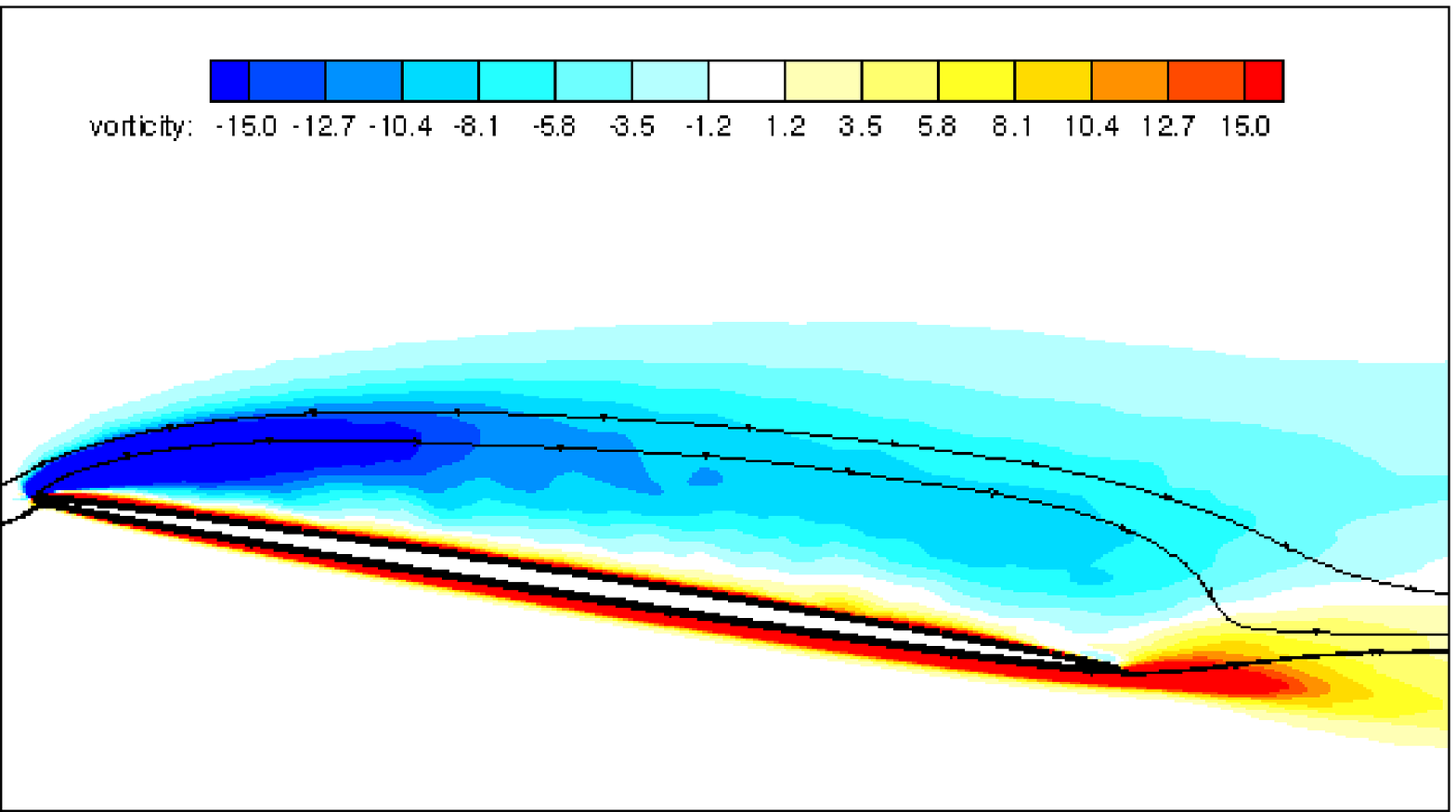}
}
\subfigure[ $\alpha=13^{\circ}$]{
\includegraphics[width=0.3\textwidth]{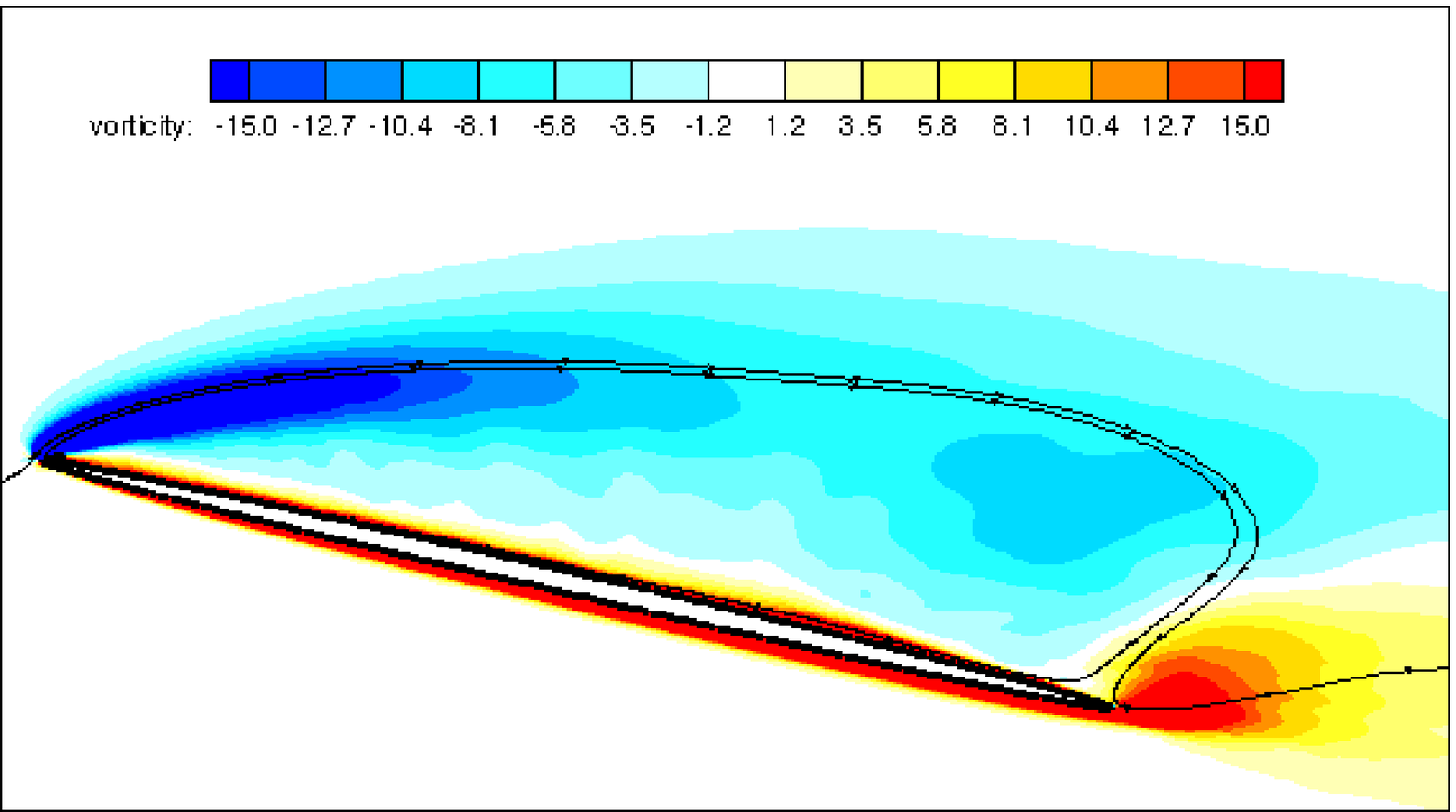}
}
\subfigure[ $\alpha=18^{\circ}$]{
\includegraphics[width=0.3\textwidth]{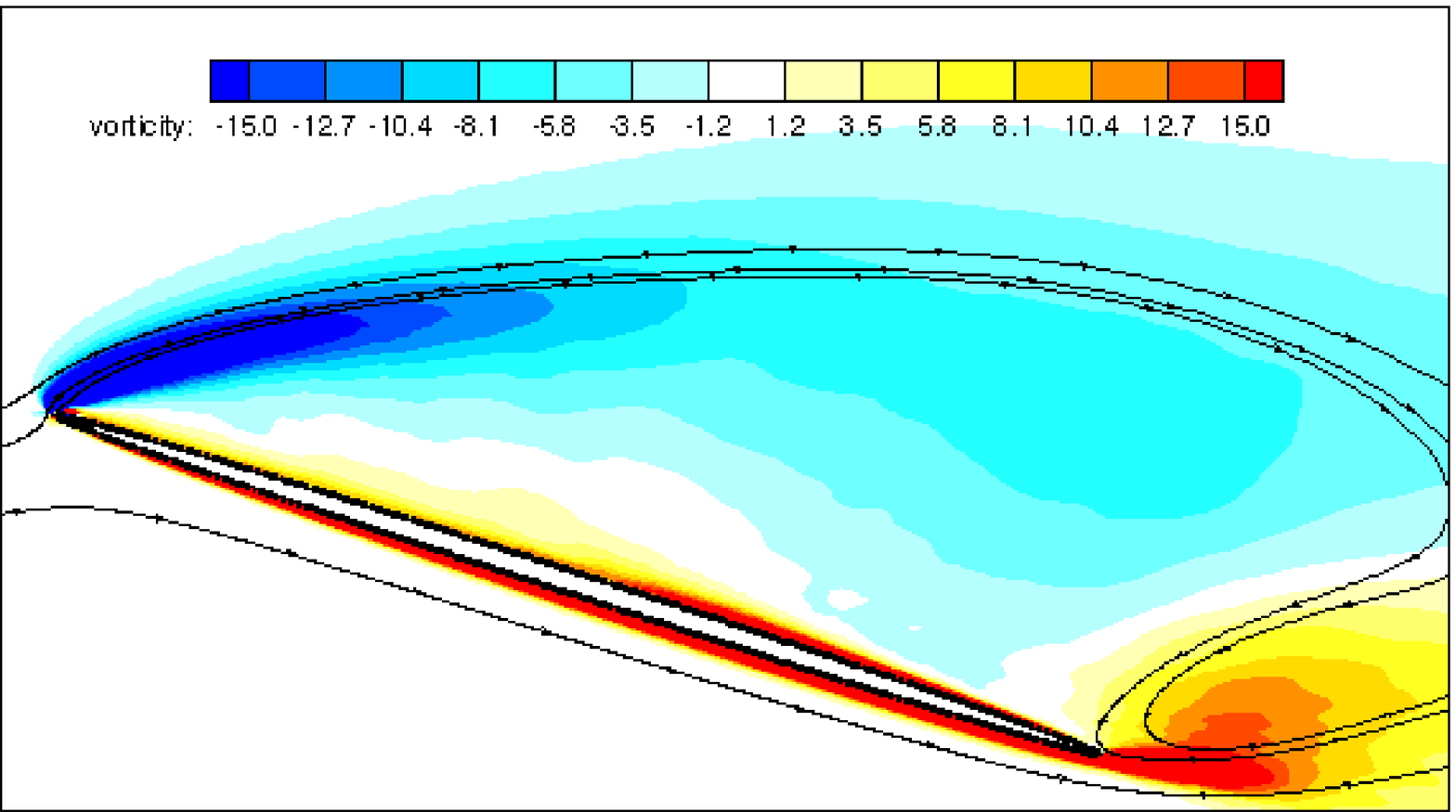}
}
\subfigure[ $\alpha=21^{\circ}$]{
\includegraphics[width=0.3\textwidth]{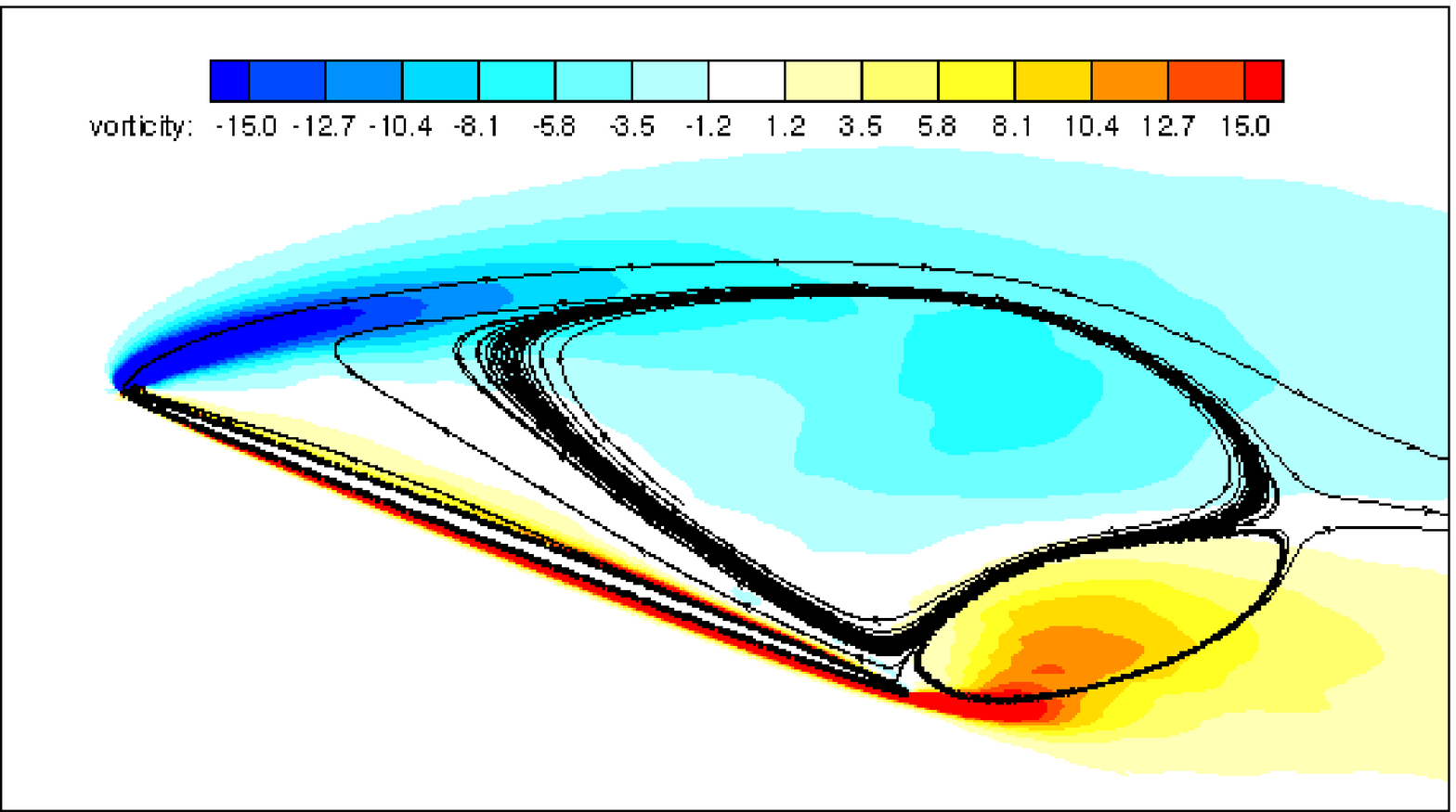}
}
 \caption{The time-averaged vorticity fields for the plate in freestream flow at fixed angles of attack.}
 \label{f:avgstatic}
\end{figure}

Classic thin airfoil theory is often used for quasi-steady predictions of lift coefficients as an airfoil changes its relative angles of attack, however this theory is formulated with potential flow approximations that assume an attached flow. In this case, it is apparent that the boundary layer is separating at angles of attack as low as $\alpha=5^{\circ}$, and since the static behavior of the airfoil is well documented, an equation of the form

\begin{equation}
\label{e:fit}
C_l(\alpha) = b_1[\log(\cosh(\alpha/b_2)) - \log(\cosh(\alpha/b_2-b_3))] 
\end{equation}

\noindent
is fitted to the computed lift coefficients in figure \ref{f:staticaoa}, where $\alpha$ is in degrees. Using a least squares fitting algorithm, coefficients of $b_1=204.4$, $b_2=9.708$, and $b_3=5.00\times10^3$ are found to best represent the static behavior of the thin-ellipse airfoil at $Re=40,000$, from $\alpha=-21^{\circ}$ to $21^{\circ}$, with the results plotted in figure \ref{f:staticaoa}.  The fitted curve from eqn. \ref{e:fit} can be used to give a more accurate quasi-steady prediction of the lift coefficient $C_l(t)$ for the heaving plate as it progresses through various relative angles of attack. 

In order to further characterize the static flow, time-averaged vorticity fields and streamlines are shown in figure \ref{f:avgstatic} for angles of attack at 5, 7, 9, 13, 18, and 21 degrees.  Due to the slender geometry of the high-aspect ratio ellipse, separation is always initiated at the leading edge.  As is typical with thin-airfoil stall, a separation bubble forms at the leading edge and the reattachment point moves aft with increasing angle of attack. The streamlines indicate a small separation bubble at 10\% of the chord for $\alpha=5^{\circ}$, consistent with the small deviation from the classic airfoil prediction in figure \ref{f:staticaoa}. At $\alpha=7^{\circ}$, the separation bubble grows to approximately 30\% of the chord, and by $\alpha=9^{\circ}$, the boundary layer separates completely.  As the angle of attack is increased beyond the onset of stall, the separated shear layer is deflected further away from the airfoil, and the trailing edge vortex becomes stronger.  To complement the mean vorticity fields, the pressure coefficient at the surface is shown for $\alpha=5^{\circ},7^{\circ},9^{\circ}$ and $13^{\circ}$ in figure \ref{f:cp}(a), which shows the progression of the steady state flow separation on the suction side of the plate.

Spectra (not shown) from the unsteady lift vs time are consistent with the time-averaged vorticity fields, showing no dominant frequencies from unsteady vortex structures for $\alpha <=5^{\circ}$.  Beginning at $\alpha=7^{\circ}$, there are peaks at the expected bluff body shedding frequency of $f c \sin(\alpha)/U_{\infty}\approx0.16-0.2$, but also strong peaks at the subharmonic in the range of $0.08-0.1$. The subharmonic peaks that are prominent at higher angles of attack are likely due to the leading edge vortices coalescing into larger vortical structures that heavily influence the instantaneous lift.

\section{Heaving Plate Simulations}
The heaving plate simulations are performed at angles of attack of $\alpha=1$, 5, 9, 13 and 18 degrees. The resulting time-averaged lift coefficient, $\overline{C_l}$, of each case is shown in table \ref{t:cl}. There are three regimes represented by these angles of attack.  The first regime is that which the flow remains mostly attached throughout the heaving cycle, which occurs when heaving at $\alpha=1^{\circ}$. In this regime, the mean lift from the heaving foil is approximately equal to that of the static foil and reasonably predicted by the quasi-steady estimate.  The second regime, from $\alpha=5^{\circ}$ to $13^{\circ}$, is characterized by the development and shedding of a large LEV, which produces significant lift enhancement compared to both the static and quasi-steady predictions. The net increase in average lift compared to the static foil ranges from $5.6\%$ at $\alpha=5^{\circ}$ to a peak of $17.2\%$ for $\alpha=13^{\circ}$.  Compared to the quasi-steady estimates, the average lift is between $17.9\%$ to $24.1\%$ higher for this second regime, emphasizing that the lift enhancement is due to the unsteady fluid-structure interaction and not just a change in relative angle of attack. The third regime is characterized by the heaving plate at $\alpha=18^{\circ}$, where the flow never completely reattaches due to the very high relative angles of attack throughout the heaving stroke.  In this regime a modest increase from the static case ($9.7\%$) and quasi-steady estimate ($5.3\%$) are reported.

\begin{figure}[htp]
\centering  
\includegraphics[width=5in]{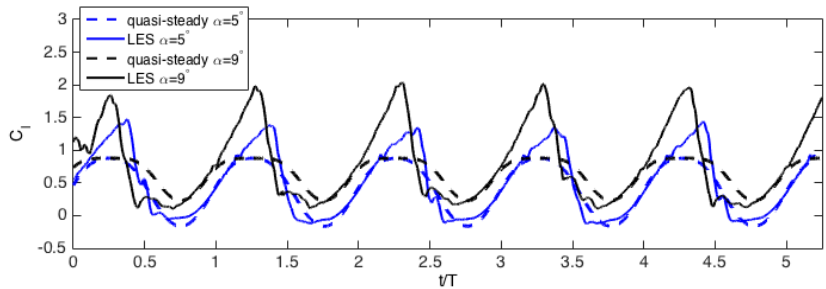}
\caption{The lift cycle for all 5 heaving cycles for $\alpha=5^{\circ}$ and $\alpha=9^{\circ}$. The transient behaivor is mostly gone after the first cycle.  Also for comparison is the predicted lift from a quasi-steady approximation.}
\label{f:heavingcycle}
\end{figure} 

An example of the five computed heaving cycles is shown in figure \ref{f:heavingcycle} for $\alpha=5^{\circ}$ and $\alpha=9^{\circ}$. During the first cycle the flow is adjusting to the acceleration terms added to the flow, but by the second heaving cycle the flow has reached a steady state heaving mode, and is surprisingly repeatable from cycle to cycle. In dashed lines is the quasi-steady prediction for each angle of attack, which peaks at $C_l \approx 0.9$, or when stall occurs for the static flow. The lift coefficient of the heaving plate surpasses the quasi-steady prediction in the first half of each stroke (downstroke) and follows the quasi-steady prediction during the upstroke.

\begin{figure}[htp]
\includegraphics[width=0.89\textwidth]{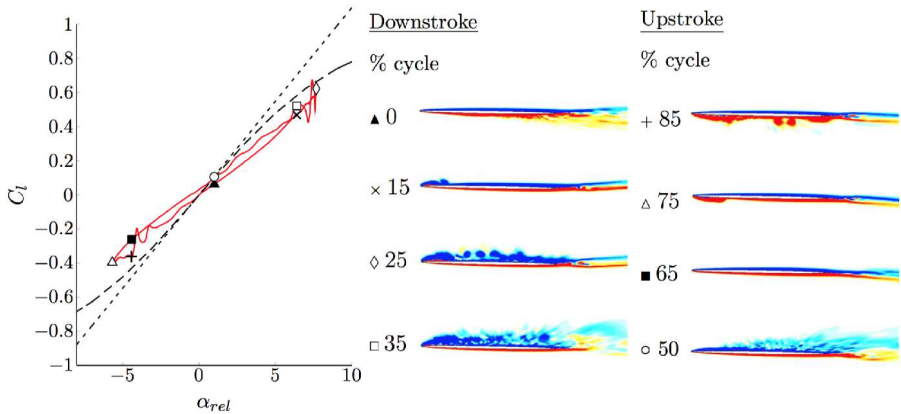}
  \caption{Heaving at $\alpha=1^{\circ}$: the phase-averaged lift coefficient versus relative angle of attack (solid) plotted against the static lift coefficients (dashed) and thin-airfoil theory (dotted). At a low angle of attack the heaving motion follows the quasi-steady prediction.  Percent of heaving cycle: {\scriptsize $\blacktriangle$} 0\%; {\scriptsize $\times$} 15\%; {\scriptsize $\lozenge$}  25\%; {\tiny $\square$} 35\%; $\circ$ 50\%; {\tiny $\blacksquare$}  65\%; {\scriptsize $\vartriangle$} 75\%; {\scriptsize $+$} 85\%. Phase-averaged spanwise vorticity, contour levels: -10 (blue) to 10 (red).}
\label{f:aoa1all}
\end{figure}

\begin{figure}[htp]
\subfigure[Heaving at $\alpha=5^{\circ}$]{
\includegraphics[width=0.89\textwidth]{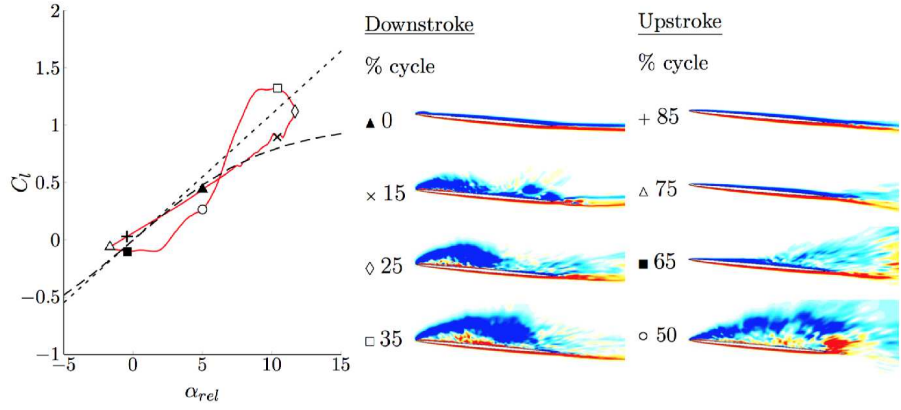}
\label{f:aoa5alla}}
\subfigure[Heaving at $\alpha=9^{\circ}$]{
\includegraphics[width=0.89\textwidth]{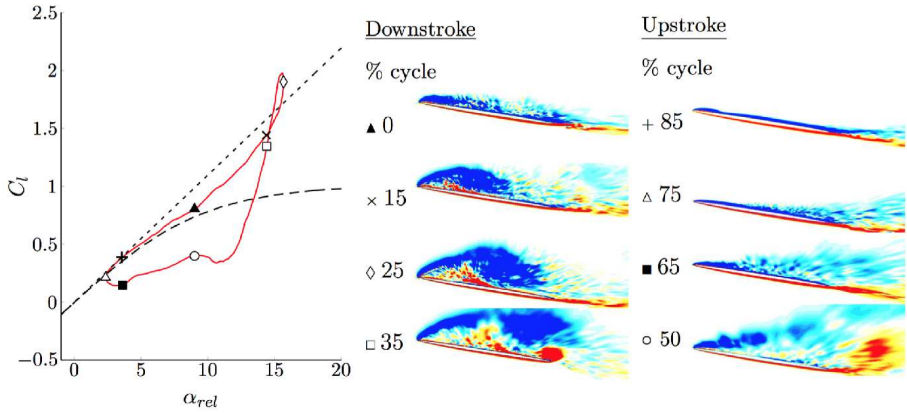}
\label{f:aoa5allb}}
\subfigure[Heaving at $\alpha=13^{\circ}$]{
\includegraphics[width=0.89\textwidth]{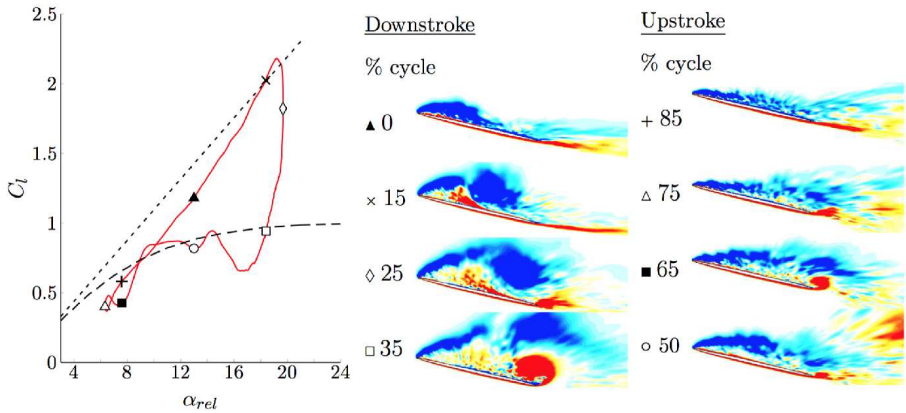}
\label{f:aoa5allc}}
  \caption{Heaving at moderate angles of attack from $\alpha=5^{\circ}$ to $\alpha=9^{\circ}$. Same legend as fig. \ref{f:aoa1all}. This regime is characterized by a significant LEV that enhances lift throughout much of the downstroke.}
\label{f:aoa5all}
\end{figure}

\begin{figure}[htp]
\includegraphics[width=0.89\textwidth]{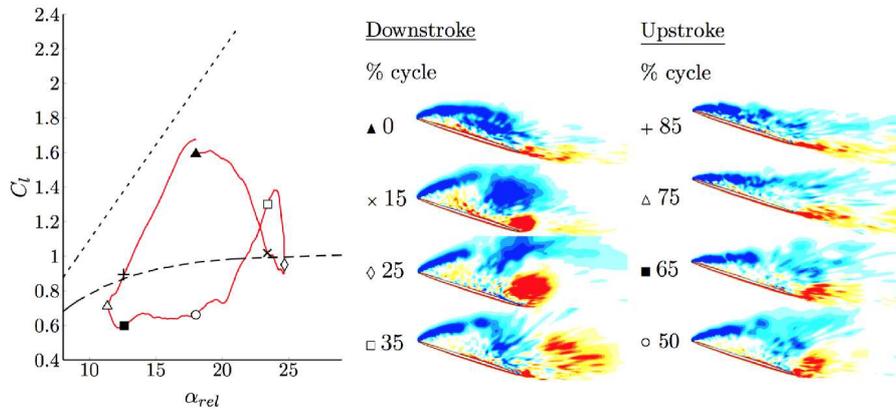}
  \caption{Heaving at $\alpha=18^{\circ}$. Same legend as fig. \ref{f:aoa1all}. At high angles of attack the boundary layer never reattaches but still enhances lift with a LEV at the beginning of the downstroke.}
\label{f:aoa18all}
\end{figure}

Figures \ref{f:aoa1all} through \ref{f:aoa18all} 
demonstrate these unsteady flow structures for various phases in the cycle paired with the time-dependent lift coefficient as a function of $\alpha_{rel}$.   The red line is the heaving plate simulations; thin airfoil theory (dotted) and the quasi-steady prediction (dashed) are included for comparison. At 0\% of the cycle ({\scriptsize $\blacktriangle$}) the plate is at its maximum height, beginning the downstroke. At 25\% cycle ({\scriptsize $\lozenge$}) the plate reaches maximum downward velocity and maximum $\alpha_{rel}$, completing the downstroke at 50\% ({\scriptsize $\circ$}), and reaching maximum upward velocity and minimum $\alpha_{rel}$ at 75\% cycle ({\scriptsize $\vartriangle$}).  The vorticity plots portray the downstroke (read from top to bottom) and the upstroke positions (read from bottom to top) corresponding with the symbols. Each of these figures represents data phase-averaged over a few cycles.  All runs had very repeatable phase-averaged forces after two cycles, with the exception of the $\alpha=18^{\circ}$ case where the lift signal was more unpredictable and thus was phase-averaged over 7 cycles. 

%

\begin{table}[htp]
  \caption{Mean Lift Coefficient, $\overline{C_l}$}
\begin{tabular*}{0.99\textwidth}{l @{\extracolsep{\fill}} ccccc}
\hline
& \multicolumn{5}{c}{$\alpha$}\\
 & $1^{\circ}$ & $5^{\circ}$ & $9^{\circ}$ & $13^{\circ}$ & $18^{\circ}$ \\ 
 \hline
static airfoil & 0.093 & 0.468 & 0.739  & 0.887 & 0.911\\
quasi-steady prediction (eqn. \ref{e:fit}) & 0.087 & 0.406 & 0.666 & 0.838 & 0.949\\
heaving airfoil & 0.093  & 0.494 & 0.785  & 1.04 & 1.00 \\
\% increase from static & 0 & 5.6 & 6.2 & 17.2 & 9.7 \\
\% increase from quasi-steady & 6.9 & 21.7 & 17.9 & 24.1 & 5.3\\
\hline
\end{tabular*}
\label{t:cl}
\end{table}

\hspace{-1in}

\begin{figure}[htp]
\includegraphics[width=0.99\textwidth]{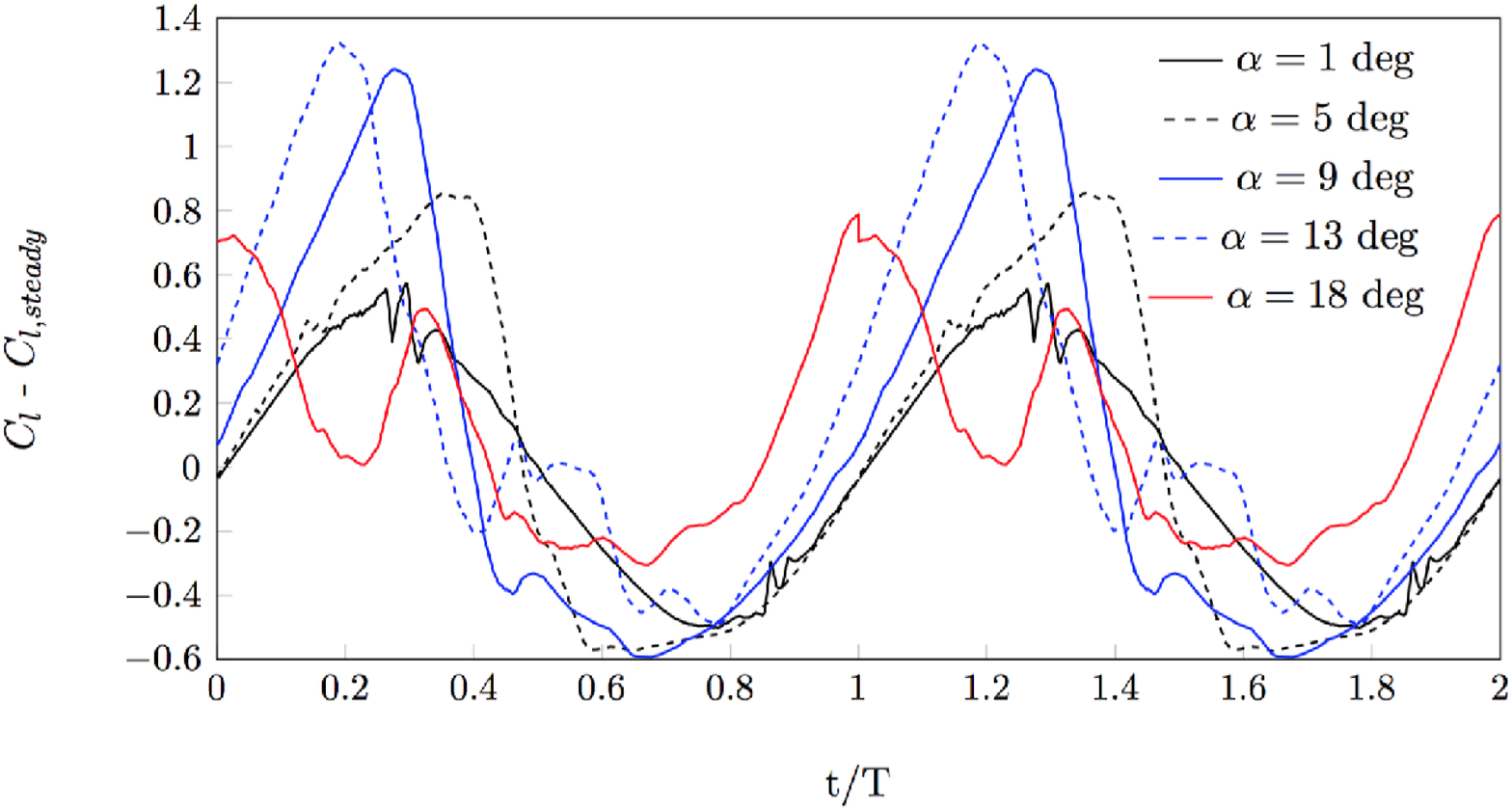}
\caption{Phase-averaged lift coefficient with the steady lift coefficient subtracted for better comparison. Two heaving cycles are shown to better display the trend that occurs upon the transition of one cycle to the next. 
}
\label{f:phaselift}
\end{figure} 

The heaving flow at an angle of attack of $1^{\circ}$ follows a sinusoidal pattern very close to the values predicted by the quasi-steady calculations with only a slight deviation at the highest and lowest relative angle of attacks (Fig. \ref{f:aoa1all}).  Such behavior is to be expected since the relative angle of attack remains below the onset of static stall at $\alpha=9^{\circ}$. The phase-averaged vorticity in figure \ref{f:aoa1all} show no prominent large-scale vortex shedding, although there is an instability that arises in the upper boundary layer during the downstroke which could be a preliminary sign of separation. The instability is composed of small vortical structures that form at the leading edge of the plate. At 15\% of the cycle these structures are only present on the front 10\% of the plate, but as the plate progresses from 15\% to 35\% these small structures continue to form then advect down the length of the plate. At the end of the downstroke these small vortices have stopped forming and previously shed vortices have merged into a turbulent boundary layer on the suction side of the plate. As soon as the plate reverses its heave direction the boundary layer quickly relaminarizes and a similar instability forms on the lower side of the plate during upstroke. Evidence of the instability can be seen in the $\alpha_{rel}$ vs. $C_l$ plot in figure \ref{f:aoa1all}.

Figure \ref{f:aoa5all} illustrates the lift behavior throughout the second regime, where a LEV dominates throughout the downstroke.  Compared to the quasi-steady estimate, the lift is significantly enhanced during the formation of the LEV on the upper surface of the plate.  The value of the maximum lift correlates well with the thin-airfoil theory approximations for $\alpha=5^{\circ}$, $9^{\circ}$, and $13^{\circ}$.  However it should be emphasized that the thin-airfoil predictions are not properly modeling the physics of the problem since they are based on fully attached flow and that they are only shown for reference.  

No significant LEV is formed when heaving at $\alpha \ge 1^{\circ}$, but by $\alpha=5^{\circ}$ there is a large unsteady vortex on the top (suction) side of the foil, which grows larger when heaving at $\alpha=9$ and reaches its peak size and strength at $\alpha=13^{\circ}$. The maximum lift peak in each case corresponds to the maximum size of the LEV before the formation of a coherent trailing edge vortex (TEV).  Once the TEV is formed, the lift decreases rapidly and the LEV is subsequently shed from the foil. Although this behavior is consistent throughout the range of angles of attack tested, the peak lift is shifted to an earlier point in the cycle as the angle of attack increases.  This is illustrated by Fig \ref{f:phaselift} in which the unsteady lift as a function of cycle time is plotted for each of the cases, over two complete cycles.  At $\alpha=1^{\circ}$ the unsteady lift is almost perfectly sinusoidal, well represented by the quasi-steady approximation.  For greater angles of attack the peak lift occurs earlier in the cycle from 40\% of the cycle for $\alpha=5^{\circ}$, which is close to midstroke, to 30\% for $\alpha=9^{\circ}$ and 20\% for $\alpha=13^{\circ}$.  The peaks also grow in strength, peaking at $\alpha=13^{\circ}$.  

At $\alpha=18^{\circ}$ the mean angle of attack is so great that the peak lift occurs at the top of the downstroke after which the LEV immediately separates and forms a separated shear layer. The shear layer never reattaches fully but slowly organizes into a vortex during the upstroke.  This vortex offers some lift benefits, but since the flow never completely attaches to the surface there are minimal lift enhancement benefits for this high angle of attack, and the mean lift is converging closer to the mean lift of the static foil.

\section{Leading Edge Vortex Dynamics}

\begin{figure}[htp]
\centering
\includegraphics[width=0.65\textwidth]{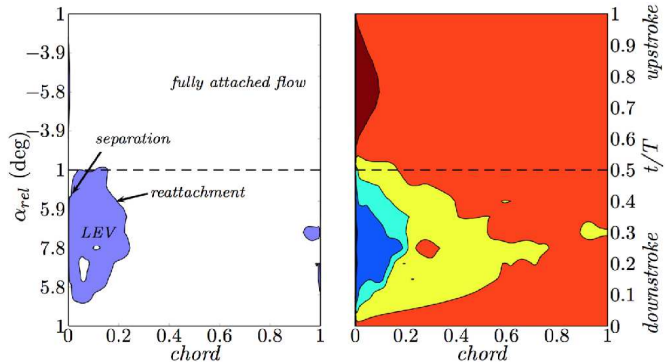}
\caption{The phase-averaged progression of flow structure formation and separation for heaving at $\alpha=1$ deg based on vorticity (left) and pressure (right) on the top surface of the foil as a function of chord length and time within a cycle. Thresholds of positive (shaded) and negative (white) vorticity  indicate flow separation and reattachment, and can be correlated with low pressure regions (contour levels from $p/\rho U_{\infty}^2$=-3.5 (blue) to 0.5 (red)). At $\alpha=1$ deg we see a small LEV associated with a low pressure region over 20\% of the chord during the downstroke.}
\label{f:xvst1}
\end{figure}

The process of boundary layer separation and reattachment on the surface of the plate can be represented by changes in the sign of vorticity, from negative when the flow is fully attached on the top surface to positive when it has become separated. Furthermore, as a consequence of separation there are significant changes to the local pressure field on the top surface, dynamically changing the lift coefficient throughout the cycle. To illustrate the connection between boundary separation and pressure throughout the heaving cycle, figure \ref{f:xvst1} displays the phase-averaged vorticity and pressure contours for the heaving plate at $\alpha=1^{\circ}$ as a function of chord location on the horizontal axis and $\alpha_{rel}$ on the vertical axis. The vorticity contours are shaded for positive vorticity representing separated flow regions, whereas the white background signifies attached flow on the top surface. 

The flow physics of the heaving plate at $\alpha=1^{\circ}$ can be explained in \ref{f:xvst1} by a small and attached LEV that covers up to $22\%$ of the chord during the downstroke, but reattaches during the upstroke for fully attached flow along the top surface. The pressure contours of the right frame of figure \ref{f:xvst1} demonstrate the region of low pressure that correlates strongly with the presence of the LEV, and is responsible for the lift enhancement.  This is consistent with the phase-averaged vorticity contours in figure \ref{f:aoa1all}. The upstroke region has a small region of high pressure at the leading edge, which is appropriate for the negative values of $\alpha_{rel}$ and the corresponding negative lift for this phase of the stroke.

\begin{figure}[htp]
\centering
\subfigure[Heaving at $\alpha=5$]{
\includegraphics[width=0.65\textwidth]{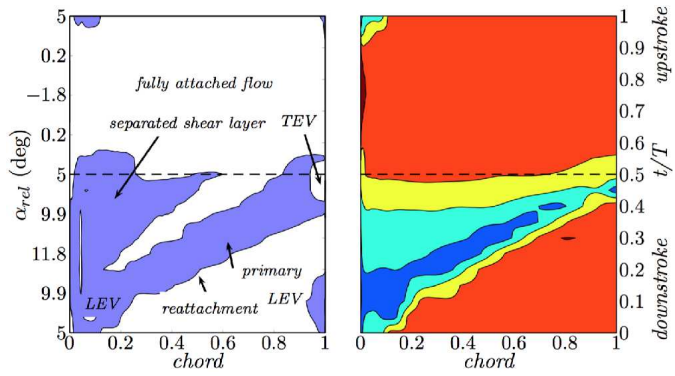}
\label{f:xvst5}
}
\subfigure[Heaving at $\alpha=9$]{
\includegraphics[width=0.65\textwidth]{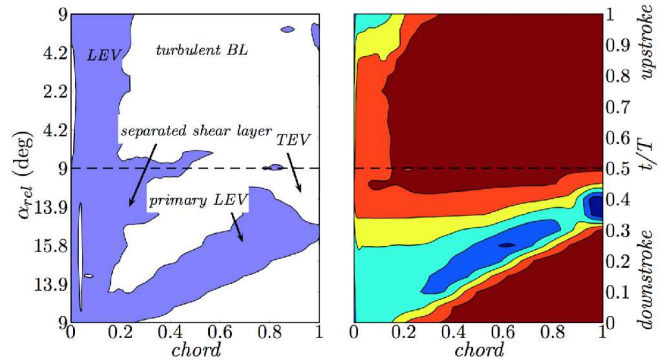}
\label{f:xvst9}
}
\subfigure[Heaving at $\alpha=13$]{
\includegraphics[width=0.65\textwidth]{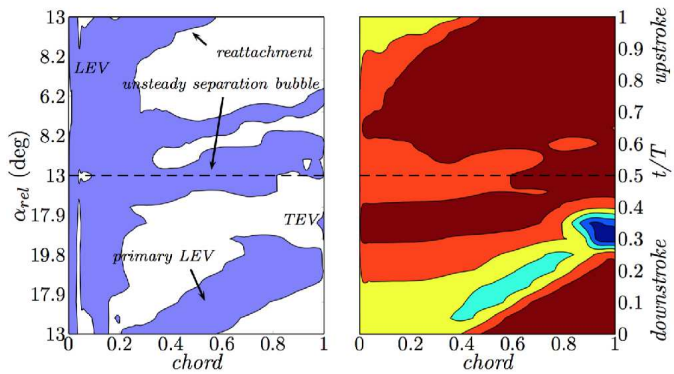}
\label{f:xvst13}
}
\caption{The phase-averaged progression of flow structures and pressure for heaving at $\alpha=5^{\circ}$, $9^{\circ}$, and $13^{\circ}$ (see legend in Fig. \ref{f:xvst1}. In each case an LEV is formed and moves from downstream creating a trail of low pressure along the chord during the downstroke.}
\label{f:xvst5913}
\end{figure} 

Figure \ref{f:xvst5} shows the progression of the flow structures leading to the lift enhancement for the heaving plate at $\alpha=5^{\circ}$. The regions of separated flow shown by the vorticity contours correlate with the regions of low pressure on the top surface of the plate.  In contrast to heaving at $\alpha=1^{\circ}$, an instability in the upper boundary layer develops rapidly at the onset of the downstroke and immediately forms into a separation bubble on the upper surface of the plate initiated at the leading edge. The flow separation at the leading edge organizes itself into a strong LEV from $t/T=0-0.2$ decreasing the pressure on the suction side of the foil over approximately $20\%$ of the chord. Instead of reattaching, this primary LEV traverses the length of the plate locally decreasing the surface pressure along its path.  As it convects downstream it leaves behind a separated shear layer that forms a secondary, weaker separation region at the leading edge associated with a weaker low pressure region.  Both the primary LEV and secondary shear layer exist and contribute to the peak value of lift which occurs at $t/T=0.4$, well after the maximum relative angle of attack at $t/T=0.25$.  The primary LEV eventually separates from the surface when a small trailing edge vortex (TEV) develops on the last 10\% of the chord.  The TEV contributes slightly to the lift profile from $t/T=0.4-0.5$.  Beginning at $t/T=0.5$ the plate begins its upstroke and all vortices are quickly convected downstream from the plate. The separated boundary layer begins reattachment from the the front to the rear during which the lift drops below that of the static plate until it completely reattaches at approximately $t/T=0.65$.

The heave stroke at an angle of attack of $9^{\circ}$ in figure \ref{f:xvst9} shows similar trends to figure \ref{f:xvst5} except that the flow never completely reattaches during the upstroke.  Thus, there is a region of separated flow at the leading edge of the plate during both the upstroke and downstroke.  Due to this separated region, as soon as the relative angle of attack begins increasing at $t/T=0.75$ (last portion of the upstroke) a LEV begins to organize and grow in strength, decreasing pressure over the first quarter chord. At the beginning of the downstroke it begins convecting downstream, remaining attached to the plate and growing in strength until $t/T=0.25$ when it reaches its maximum relative angle of attack and just before its maximum lift at $t/T=0.3$. Just before the reattachment line reaches the trailing edge the primary vortex separates from the plate. Taking its place on the aft $10\%$ of the chord is a TEV which locally decreases pressure for the remainder of the downstroke. The separated shear layer from the leading edge begins its reattachment process during the upstroke but the relative angles of attack are too large to completely reattach and leave behind the small separated region on the leading edge which will form into a LEV on the next cycle.

Vorticity and pressure contours for the heaving plate at $\alpha=13^{\circ}$ are shown in figure \ref{f:xvst13}.  It follows similar trends from figure \ref{f:xvst9} except that the primary LEV begins its development slightly earlier in the upstroke, and proceeds to grow and convect at a faster rate. As is also seen in figure \ref{f:phaselift}, the peak lift occurs at $t/T=0.2$, which is early in the downstroke cycle. The shift in the peak lift as the angle of attack increases can be explained by the earlier onset of LEV formation that occur at such high relative angles of attack. The LEV pinches off at approximately $t/T=0.3$ and leaves behind a highly separated shear layer emanating from the leading edge. The shear layer attempts to reattach and forms a large unsteady separation bubble over the length of the plate from $t/T=0.4-0.65$, which gives rise to a secondary peak in the lift curve in figure \ref{f:phaselift}. The separated region diminishes in size throughout the upstroke, and by $t/T=0.75$ it has transitioned into a turbulent boundary layer with a very thin separation bubble at the leading edge. This separation bubble will give rise to the primary LEV of the next heave cycle.

\begin{figure}[htp]
\centering
\includegraphics[width=0.65\textwidth]{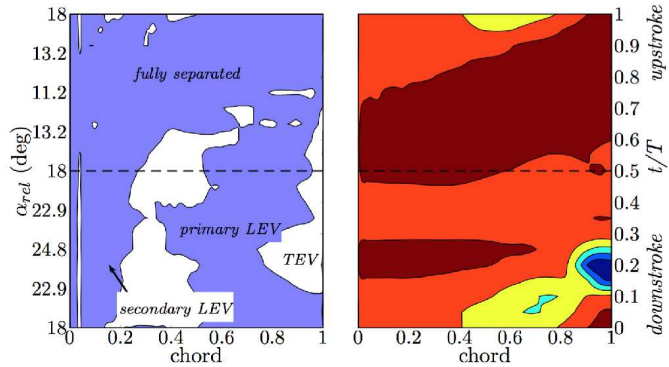}
\caption{The phase-averaged progression of flow structures and pressure for heaving at $\alpha=18^{\circ}$ (see legend in Fig. \ref{f:xvst1}), in which case the flow is completely separated from the surface. The presence of the LEV and TEV, and their contributions to the lift enhancement are shown through phase-averaged vorticity and pressure contours.}
\label{f:xvst18}
\end{figure} 
 
The heaving plate at $\alpha=18^{\circ}$ is the third regime of the flow in which the boundary layer remains fully separated for the duration of the stroke. Relative angles of attack for this case range from $11.2^{\circ}$ to $24.8^{\circ}$, which are both past the static stall angle. Since the boundary layer never reattaches, the shear layer forms a recirculation region by reattaching at the trailing edge as the relative angle of attack increases. The unsteady separation bubble becomes more coherent and forms a single large primary vortex over the entire chord at the end of the upstroke. This vortex is then shed, along with a trailing TEV, but the shear layer quickly organizes to form a second, weaker, vortex on the upper surface of the plate in the same manner as the first. Each of these recirculation regions is highly unsteady and weaker than than the coherent vortices displayed at lower angles of attack, resulting in a more unpredictable lift shown in figure \ref{f:aoa18all} and a significant decrease in the peak values of lift compared to the other angles of attack in figure \ref{f:phaselift}.  Although the lift enhancement has decreased significantly at $\alpha=18^{\circ}$, there is still 9.7\% increase from the static (fully stalled) plate, and a 5.3\% above what is predicted from quasi-steady analysis. This indicates that even a weak and highly unsteady reattachment of the boundary layer can be beneficial in terms of increasing the mean lift.

\section{Conclusions}

Simulations of a heaving flat plate at fixed angles of attack of $1^{\circ}$, $5^{\circ}$, $9^{\circ}$, $13^{\circ}$, and $18^{\circ}$ are simulated and the mean and phase-averaged lift was calculated and compared with the resulting flow fields. The investigation was designed to complement the experimental work of a self-excited flapper \cite{Curet2013}, whose lift increased upon transition from a steady to flapping mode. The heaving plate model simulated is to capture the two-dimensional lift-enhancement effects that stem from leading edge separation at a transitional Reynolds number of 40,000. 

The results from the heaving plate simulations are characterized by three regimes, a fully attached regime ($\alpha=1^{\circ}$), a fully separated regime ($\alpha \ge 18^{\circ}$), and a transitional regime ($\alpha=5^{\circ}-13^{\circ}$) whose instantaneous and mean lift benefits greatly from the presence of LEV and other unsteady flow structures that develop due to the heaving motion.

Heaving at an angle of attack of $1^{\circ}$, which is representative of the first regime, the lift is $6.9\%$ greater than that predicted by a quasi-steady model. However the general trends are well predicted by the model since the relative angle of attack remains smaller than the static stall value and only minimal boundary layer separation is observed.  

In the transitional regime, represented by heaving at angles of attack between $5^{\circ}$ and $13^{\circ}$, the lift enhancement was between $18-24\%$ greater than the quasi-steady predictions and $5-17\%$ greater than the corresponding static flow. Since static stall begins at $9^{\circ}$, each of these plunging flows extends well into the static stall region but is able to maintain lift due to the presence of coherent leading and trailing edge vortices, and also due to less coherent unsteady separation bubbles.  The lift enhancement is due to a significant decrease in pressure on the suction side of the foil during the downstroke, although the exact time at which the lift peaks within the downstroke is earlier with increasing angle of attack since the primary LEV forms .  At $\alpha=5^{\circ}$ the lift peaks at $t/T=0.4$ and at $\alpha=18^{\circ}$ the lift peaks at $t/T=0$.  The maximum lift coefficient increases with angle of attack until it peaks at 2.1 for $\alpha=13^{\circ}$, which has a mean value of 1.04, 17\% greater than the steady state flow at $\alpha=13^{\circ}$.

A detailed analysis of the vortex formation combined with the pressure contours along the surface over a complete cycle offer a more complete view of the mechanisms for lift enhancement.  The boundary layer separates at the leading edge with increasing relative angle of attack but forms into a coherent vortex which keeps the flow loosely attached to the top surface and decreases the pressure on the suction side of the plate.  Coherent vortices are convected down the chord reaching a peak lift when they are at their maximum size, and aided by the presence of a TEV that contributes to the low pressure on the suction side of the foil. Shortly after the TEV is formed, the LEV sheds and lift significantly drops.  The resulting separated shear layer either completely reattaches ($\alpha \le 5^{\circ}$), partially reattaches into a laminar separation bubble ($9^{\circ} \leq \alpha \leq 13^{\circ}$), or forms a turbulent region of separated flow ($\alpha = 18^{\circ}$) during the upstroke. 

\section{Acknowledgements}

This work was performed under a NSF Postdoctoral Fellowship, Award No. DBI-0906051. Computational resources were provided by DoD HPC through AFOSR grant 302102 under Dr. Douglas Smith and through resources and services at the Center for Computation and Visualization at Brown University.

\section*{References}

\bibliography{mybibfile.bib}

\end{document}